\documentclass[runningheads,a4paper]{llncs}
\usepackage{makeidx}  
\usepackage{amssymb}
\usepackage{graphicx}
\usepackage{url}
\newcommand{\keywords}[1]{\par\addvspace\baselineskip
\noindent\keywordname\enspace\ignorespaces#1}
\makeatletter
\def\captionof#1#2{{\def\@captype{#1}#2}}
\makeatother
\begin{document}
\mainmatter              
\title{The Crypto-democracy and the Trustworthy}
\titlerunning{The Crypto-democracy and the Trustworthy}  
%
\author{S\'ebastien Gambs$^{1}$ \and Samuel Ranellucci$^{2}$ \and Alain Tapp$^{3}$}
\authorrunning{Gambs, S., Ranellucci, S. and Tapp, A.}   
\institute{$^{1}$Universit\'e de Rennes 1 - INRIA / IRISA, France\\
$^{2}$Department of Computer Science, Aarhus University, Denmark\\
$^{3}$DIRO, Universit\'e de Montr\'eal, Canada\\
\{\url{sgambs@irisa.fr},\url{samuel@cs.au.dk},\url{tappa@iro.umontreal.ca}\}
}

\toctitle{The crypto-democracy and the Trustworthy}
\tocauthor{S\'ebastien Gambs, Samuel Ranellucci, Alain Tapp}

\maketitle              

\begin{abstract} 
In the current architecture of the Internet, there is a strong asymmetry in terms of power between the entities that gather and process personal data (\emph{e.g.}, major Internet companies, telecom operators, cloud providers, \ldots) and the individuals from which this personal data is issued. In particular, individuals have no choice but to blindly trust that these entities will respect their privacy and protect their personal data. In this position paper, we address this issue by proposing an utopian crypto-democracy model based on existing scientific achievements from the field of cryptography. More precisely, our main objective is to show that cryptographic primitives, including in particular secure multiparty computation, offer a practical solution to protect privacy while minimizing the trust assumptions. In the crypto-democracy envisioned, individuals do not have to trust a single physical entity with their personal data but rather their data is distributed among several institutions. Together these institutions form a virtual entity called the Trustworthy that is responsible for the storage of this data but which can also compute on it (provided first that all the institutions agree on this). Finally, we also propose a realistic proof-of-concept of the Trustworthy, in which the roles of institutions are played by universities. This proof-of-concept would have an important impact in demonstrating the possibilities offered by the crypto-democracy paradigm. 
\end{abstract}

\section{Introduction}

The recent revelations from Snowden about the NSA's ability to eavesdrop on communications, as well as to track the digital traces left by Internet users, clearly demonstrates that we are currently moving towards an information age that is not so far from 1984. Since the original novel from Orwell, the concept of Big Brother has become a powerful meme. Indeed even if few people understand the subtleties of modern cryptography, the potential abuses resulting from the massive collection of personal data and the associated invasion of privacy is very present in people's mind. Thus, Big Brother is now synonymous of a dystopian future in literature, movies, arts and general culture.

The amount of information collected by Internet companies and third parties on individuals increases every day, both in quantity and in diversity. For instance, some actors have access to personal data such as social relationships, email content, income information, medical records, credit card and fidelity card usage, pictures taken through public and private cameras, personal files, navigation behavior, location, biometrics, data issued from quantified self, \ldots, just to name a few. In addition, the ability to capture and record all aspects of life (both real and virtual) of users has increased dramatically recently due to new technological developments. For instance, Memoto\footnote{\url{https://www.kickstarter.com/projects/martinkallstrom/memoto-lifelogging-camera}} was originally a Kickstarter project whose objective was the development of a small camera worn by the user that would automatically generate a picture of its surroundings that would be associated with GPS position and the corresponding timestamp. Even more recently, the concept of Google glass has appeared to be very attractive to some users. However, its adoption will also certainly exacerbate the privacy concern of others. 

On one hand, this massive collection of information raises many privacy issues since most data is personal and thus sensitive by nature. Already today, the accumulated amount of data collected is very significant but the possible abuses are still limited because its potential cannot be fully exploited, both for legal and technical reasons. For instance, the possibility of cross-referencing databases is often limited due to privacy laws regulating the gathering and processing of personal data. In addition, it is expected that the inference capabilities will increase dramatically as more and more data become available and gets concentrated in the hands of a few actors. Thus, we can reasonably believe that we do not have yet a full-fledged Big Brother because all this information is not accessible easily by one entity (except maybe the NSA).

On the other hand, the information captured could also be used in a number of useful and innovative manners. For instance, currently in Russia many cars are equipped with cameras in order to prevent corrupt policemen from charging a driver with a fictitious crime or to simplify insurance settlements. Thus, the record of this information can act as a safeguard against corruption. Coming from the quantified self movement, it has been suggested to rely on devices that regularly study your body while you shower to detect diseases and important changes such as pregnancy. Aggregate information also allows to pinpoint infections and cancers as well as their sources, such as poisoned wells.

With the advent of Big Data, machines will be able to perform fine-grained inferences that are not yet possible. While the basic statistical (actuarial) model is already at least as good (and often better) than experts at making predictions \cite{BT02}, machine learning has the potential for extracting even more useful information from large amount of data. This information can be used for good (\emph{e.g.}, finding a link between some profiles and particular sicknesses) or for bad (\emph{e.g.}, denying to someone the access to an health insurance because he is classified as a high risk profile). Ultimately, the machine might even become better at predicting the behavior of an individual than the person himself. For instance, a teenager has recently experienced the strange situation in which her father learned that she was pregnant due to a targeted advertisement that she received in her mailbox based on the profile constructed by the supermarket company out of her purchasing list\footnote{\url{http://www.nytimes.com/2012/02/19/magazine/shopping-habits.html?pagewanted=1&_r=2&hp&}}.

One possibility for protecting privacy could be to work on anonymous or pseudonymous data. Unfortunately, anonymizing data in a sound and robust manner is a very difficult and sometimes even impossible task \cite{Ohm12}. For instance, simply removing the personally identifiable information from the released data is usually insufficient to protect the privacy of a user. Indeed, it is possible that the combination of some attributes, which individually are innocuous, could act as a quasi-identifier and thus be used to de-anonymize the data. Thus, there is always the risk that an anonymous profile can be re-identified and linked to a real identity. In addition, if the data is composed of the queries or the mobility traces of a user, then this data is so rich that it can be used to build a very detailed profile of a user containing a lot of personal information. Even more, this profile (whether it is or not anonymous) can have a tangible impact of the life of the user. For instance, this profile might impact the price paid for a product or in an extreme case a service could be denied because of this profile, which is a form of discrimination.  

Another possibility for an individual to protect his privacy would be to segregate himself from technology by not sharing information in any way. However, we believe that such an approach is both a step backward and impossible in practice as the digital traces of the actions of the user are collected often by systems that he is unaware of or he has limited or no control on. For instance, in most cities the users of a public transportation system constantly leave traces of their whereabouts through their transport pass. Thus, the main challenge is to be able to exploit the vast possibilities offered by the use of personal data in a secure and private manner. 

In this position paper, we state our position regarding the utopian application of secure multiparty computation to democracy and propose a practical project on this theme. The outline of the paper is the following. First in Section \ref{sect_trust}, we discuss how trust is a central notion to any solution that stores and processes personal data. In particular, we highlight the trust assumptions that are usually made (implicitly or explicitly) by the current existing architectures. Then in Section \ref{sect_utopia}, we introduce, what we coin as the ``crypto-democracy utopia'', in which cryptographic techniques are use to distributively implement a (virtual) trustworthy party that we coin as the Trustworthy. The role of the Trustworthy is to manage the processing and storing of personal data in a secure and private manner while minimizing the trust assumptions. Afterwords in Section \ref{sect_soa}, we review the state-of-the-art of cryptographic techniques such as secure multiparty computation and secret sharing that constitute the building blocks of our proposition for the Trustworthy before describing in Section \ref{sect_poc} a proposition for a proof-of-concept downscale version of the Trustworhty. Finally, we conclude in Section \ref{sect_conclu}.

\section{Architecture of Trust}
\label{sect_trust}

Trust is an essential and unavoidable concept strongly linked to security and privacy. However, it is also a notion that cannot be easily and universally defined. In this section, we discuss the notion of trust and we make explicit the trust assumptions done by existing architectures. This will enable us later to put in perspective the trust assumptions that we are making ourselves with our approach in the following sections. 

{\bf Making trust explicit.} In cryptography, trust is one of the fundamental element. At the same time it is often left aside and not defined explicitly. For instance, most of the fundamental results in cryptography assume that the main participants, Alice and Bob, are themselves cryptographers, that they each have access to a computer that is fully secure and under their control and that the software they execute on this computer is well known and understood by them. In addition, when involved in the protocol other assumptions that are sometimes made is that Alice and Bob perform no other task when participating to the protocol and that they do not run several sessions of the protocol in parallel. 
The adversary model has also to define precisely the capacities of the attacker, both in terms of the attacks that it can performed and the ressources he can used (\emph{e.g.}, memory, computational power, eavesdropping capability, \ldots)
Thus, in practice in order for the protocol to be \emph{really} secure, several trust assumptions have to be made.

In a different context, when using a smartphone a user has to trust that the hardware (\emph{e.g.}, the SIM card) and software installed on his device are secure and will not leak private information about him without his consent. In addition, when the user downloads an application on the App Store, the user trusts that Apple has verified the behavior and the code of the application and that there is no risk in using it. In practice, these trust assumptions can be betrayed if the smartphone is infected by malware or if information leaks out of the device through a malicious application. On the Internet, when the user gives his data to a company or a cloud provider, in some sense he is doing an act of faith that this entity will respect his privacy. Similarly, when navigating on the web, for the user to believe that a website will forget the digital traces left by his interactions if it follows the {\em Do Not Track} Initiative is also a kind of wishful thinking. 

Trust is also a central ingredient of public-key infrastructures in which certification authorities are responsible for certifying the identity of an entity or in its decentralized version of the Web of Trust (\emph{i.e.}, through PGP-like solutions). Finally, even in quantum cryptography in which the security of the protocol is supposed to depend on the physical laws of quantum mechanics, ultimately one has to trust the hardware provided by the quantum cryptography company.

{\bf More crypto does not mean more trust.} Adding more cryptography to a particular service does not necessarily make it more trustworthy. For instance, a few years ago the debit card in Canada were equipped with a secure chip and a PIN in contrast with the credit cards themselves (at that time). Thus, one can say that the debit card was more trustworthy as it relies on more advanced cryptography and clearly provide more security, but was is really the case? In reality, several persons were facing serious issues after being stolen their PIN. Indeed, the reaction of the bank was that the system in place is secure and that the users are guilty of not having protected their PIN. Those unlucky customers were having the burden of proof, mainly because the security of the system was unquestioned. In contrast, the lack of security of the credit card was such that the bank would have almost no capacity to contradict a person claiming not to have made a purchase. Of course, the total cost of the fraud for credit cards was huge and the lack of security was clearly very costly for the credit card company, and thus indirectly for their clients. Still this example illustrates that stronger cryptography does not automatically improve trust. Another common example is the situation in which a user is forced due to the password policy to choose a very complex password and to change it very frequently. In this situation, it is often the case that the user will end up writing his very secret password on a post-it next to his screen or that he will never log out. This situation also exemplifies that adding more security features can sometimes lead to less security if it is not balanced with usability. In addition, the user also has to trust that the current implementation of cryptographic primitives is indeed doing what it is supposed to do.

{\bf One should not have to trust the goodwill of others.} Another aspect of trust concerns the potential corruption of insiders. In particular, it is very dangerous to base security on trusting trust users as illustrated by the following anecdote. In this anecdote, one of the author was asked by one of his colleague why he is using a Gmail account instead of the email system provided by his department. In particular, his colleague insisted that he should not share all this private data, personal email and other sensitive information with companies such as Google. This means that the colleague was putting more trust on the technical support of the department than on Google itself. The response of the author to this criticism is very simple. On one hand, it is true that most of the staff from technical support in charge of the email accounts are very nice persons that it know the members of the department for a long time. Thus, it can be assumed that they do not to spy on the members of the department. On the other hand, one can easily imagine a scenario in which a conflict might arise between a person from the technical support and a professor, in which case the access to the email account might become problematic. In contrast, while Google also has incentives to spy on email (\emph{e.g.}, for personalized services and targeted advertising), it has no interest in leaking this data while an angry member of the technical support might be tempted to do so. One typical trust assumption is that rational players have to be honest, which unfortunately does not protect against a fully malicious player.

To summarize, while cryptography is a prerequisite to build security, in all known systems there is also a question of trust. For instance, the user might have to trust software, laws, computers, networks, companies, humans or governments. Thus, the fundamental question is always: who are you ready to trust and to which extent. In particular, while trust assumptions \emph{always} have to be made, one should aim for a solution in which these trust assumptions are minimal, realistic and consistent. In particular, we believe that the security and privacy guarantees provided by an architecture should not depend on the trust that one puts on a \emph{single} entity. In the following sections, we will discuss how cryptography can reduce the semantic gap between the security provided in theory, the one achieved in practice and the one perceived by the end user.

\section{The Crypto-democracy Utopia}
\label{sect_utopia}

We believe that society is in need of an elegant solution to answer to the ever growing privacy concerns. This solution should provide strong privacy guarantees while still bringing the benefits of the knowledge that can be extracted from the collected data. While we recognize that the suggestion of an entity, collecting and managing all private and public information, evokes right away the dystopian Big Brother, in this section we will describe an utopian alternative based on sound science. In particular, we believe that techniques originating from secure multiparty computation can reconcile the antagonist needs of privacy and the ability to exploit personal data for the benefit of individuals and the greater good. While this utopian dream is already possible in theory with existing cryptographic techniques, realizing it in practice requires a significant increase in computation power and communication efficiency as well as algorithmic advances, not to mention a deep social revolution. 
However, both technology and cryptography are improving at a fast pace. For instance during the last 50 years, the processing capacity of computers has been multiplied by a factor close to a 1000 times per decade. Thus, what seems to be out of reach of current technology might become possible a near future. In addition, new breakthroughs in cryptography have been invented on a regularly basis. For instance, the concept of fully homomorphic encryption was only materialized in 2009 \cite{Gentry09} but its efficiency has been improved rapidly under thorough efforts of the cryptography community.

{\bf The Trustworthy.} In this section, we propose a cryptographic approach to democracy, which is respectful of the privacy of citizens due to the way personal data is managed. In a nutshell, a set of carefully chosen independent institutions will implement a virtual entity, called the \emph{Trustworthy}, in a distributed and secure manner using known cryptographic techniques. The Trustworthy would be responsible for managing the public and private information related to individuals. None of the institutions would have directly access to the private data. Instead a precise access structure would specify the level of collaboration necessary between these different institutions before it is possible to extract or use data from the Trustworthy. For instance, one possible access structure could be that a majority of the institutions have to be involved before the private information considered can be retrieved from the system, while another one could be that unanimity is required if the piece of data considered is really sensitive. In addition, to having the capacity to retrieve the data, the institutions would be able to compute over it (provided that the predefined number of institutions accept to collaborate), thus creating new knowledge. Note that the users are not involved in this computation as once a user has trusted the Trustworthy with his data, the Trustworthy can perform operations on it without further interaction with the user. Another objective of those institutions is to guarantee the correctness and appropriate behavior of the equipment, software and infrastructure under their control. This is especially important for the access points in which new fresh data can be entered in the trusted third, both to avoid the possibility of introducing fake data in the system and also to ensure that the data of a user does not leak at this entrance point.

{\bf Choice of institution and access structure.} The choice of institutions that will be part of the distributed implementation of the Trustworthy is a central aspect of the system and many scenarios are possible. However, in all cases, the choice of the institutions should be done in a way reaching a \emph{quasi-consensus} in society. In practice, this does not mean that each citizen would have to trust all the institutions. On the contrary, if unanimity is required before data can be accessed then it is enough if each individual trusts that at least one of the elected institutions behaves in way that protect his privacy and civil rights. In addition, the structure chosen to reach this utopian goal is likely to be different from the ones that are currently used in most democratic countries. In particular, we believe that in order to avoid the transformation of the trusted third into Big Brother, each institution should not be under the control of the same entity (\emph{e.g.}, the government). Instead these institutions should represent in a balanced way different categories of the population and groups of interests. Examples of institutions could be a political party, a college of professionals, a religious group, the supreme court, a data protection authority or a non-profit organization, just to cite a few. Of course, the set of institutions does not need to be fixed for eternity and new institutions with a clear mandate could be elected to participate to the realization of the crypto-democracy.
  
For a given set of institutions, there are many possible designs for the access structure. In a nutshell, an access structure is a set of rules specifying the type of collaboration and consensus required from institutions to retrieve or compute on data. An access structure stating that the collaboration of at least $k$ amongst $n$ institutions is needed is both generic and powerful as it encapsulates access policies such as majority, unanimity and quorum. Of course, not all type of data should require the same threshold but we believe that the previously stated ones cover most useful situations. In addition even if not all data should be treated equally, having too many different access structures could render the system unnecessarily complex. 

{\bf Consistent, simple and self-explanatory trust model.} In the crypto-democracy paradigm, there is a strong correlation between the trust assumptions that are made in theory and how they should really be in practice. In particular, all institutions would certify that the entrance point in which data is fed into the system, including its hardware and software, is compliant with what is publicly expected to guarantee that there is no undue monitoring of the data. Furthermore, when the unanimity access structure is applied, it is clear from the user point of view that he only has to trust a single institution in being honest in order to trust the system, which is exactly in concordance with the theory. Another important aspect of trust is how robust is the system under the corruption of some institutions. Indeed, having a friend that works in an institution and is willing to provide private data is one thing, but having a corrupted friend in each institution is a conspiracy. 

{\bf Hardware, software and infrastructure requirements.} In practice, each institution would need to own a super computer or a dedicated cluster and to have a team of cryptographers and security experts. In addition, each institution should scrutinize and monitor very carefully all the materials and the employees working under their supervision, with a special care to the entrance point for fresh data. While such requirements have a non-negligible cost, we believe that trust, privacy and security are sufficiently important concepts to deserve such an investment. For instance in Montreal, there is currently the project to build a new bridge that will cost around 5 billions US dollars. While the project is fundamental for the city, this bridge is only one among others connecting the island of Montreal to the rest of Quebec. In comparison, the large data center that Google constructs in Singapore costs 120 millions US dollars and thus 5 billions would pay for 40 of them.

\section{State-of-the-art in Secure Computation}
\label{sect_soa}

In this section, we give a brief overview of the existing modern cryptographic techniques that could be used to implement the crypto-democracy. In theory at least, those techniques provide all the necessary ingredients to concretize the different aspects discussed in the previous sections. However to realize them in practice, the following issues have to be tackled: cost, speed, and scalability. After reviewing the state-of-the-art of the cryptographic primitives in this section and discussing the current existing implementations of such technologies, we describe in the next section a proposal for a downscale version of the Trustworthy that can be realistically implemented at a very reasonable cost. 

\subsection{Secure Multiparty Computation}

Informally, \emph{secure multiparty computation} \cite{Goldreich09,CD14,CDN14} (which we simply call \emph{secure computation} for the rest of this paper) is the field of cryptography studying how to securely and privately implement a distributed function that depends on the private inputs of many participants. Another way to phrase it is to say that the objective of secure computation is to emulate a trusted third party (\emph{e.g.}, the Trustworthy) through a distributed protocol between participants without actually requiring that every participant trust everyone else. 
Security is generally defined with respect to a particular adversary model, which captures the actions that can be performed by dishonest participants. For instance, some participants might be \emph{honest-but curious}, meaning that they follow the recipe of the protocol but collude together by exchanging their knowledge and information about the messages they have seen in order to break the privacy of the input of honest participants. In a stronger adversary model, participants can be \emph{fully malicious} and cheat arbitrarily. In this setting, in addition of breaking the privacy of the input of honest participants, the objective of the adversary can be to attack the correctness of the protocol (\emph{i.e.}, by influencing its output) or its robustness (\emph{i.e.}, by making it abort). Thereafter, we will use the term ``\emph{trust model}'' to encompass at the same time the adversary model considered as well as the trust assumptions that are made (both implicit and explicit).

More formally, we can define the general task of secure computation in the following manner. 
\begin{definition}[Secure computation]
With respect to a particular trust model $T$, a protocol securely computes a function $f$, if for any input $X = (x_{1},...,x_{n})$ provided by $n$ participants, as long as the trust assumptions of  $T$ are not betrayed, all participant only learn $f(X)$ and no additional information leaks from the protocol.
\end{definition} 

This definition does not prevent the function itself from revealing sensitive information regarding the private input of specific participant (this issue is magnified if several functions are computed on this private data).This problem has been thoroughly studied in the literature and approaches such as the notion of differential privacy have been proposed to mitigate this issue \cite{DMNS06,Dwork08}.

\subsection{Secret Sharing}

Secret sharing \cite{Shamir79} is a cryptographic technique that can be used to distribute a secret between several participants in such a way that only a legitimate group of participants can reconstruct the secret. For instance, in a $k$-out-of-$n$ secret sharing scheme, any set of $k$ or more participants out of $n$ can reconstruct the secret but less then $k$ participants have absolutely no information on the secret. In this way, the code of a bank safe could be split in such way that it can be reconstructed only if two out of three vice presidents join their personal shares (\emph{i.e.}, piece of the information). Most of the existing secret sharing schemes are information-theoretically secure, which means that an unauthorized group does not have in his hands the information to reconstruct the secret regardless of his computational power.

In this paper, we use secret sharing as a tool enabling users to distribute their secret across several institutions. In addition, secret sharing often forms the basis of secure computation protocols. More precisely, several protocols relies on an extension of secret sharing known as verifiable secret sharing. In standard secret sharing, if the shares are corrupted then the secret is lost and cannot be reconstructed. In contrast, Verifiable Secret sharing \cite{BSSB85} allows the reconstruction of the secret as long as a restricted amount of shares remain uncorrupted. Thus, verifiable secret sharing ensures the validity of the data.

\subsection{Secure Computation with an Honest Majority}

The two fundamental results of secure computation have appeared the end of the eighties \cite{GMW87,CCD88}. In a nutshell, these results consist in generic constructions showing that under the assumption that strictly more than two thirds of the participants are honest then \emph{any} function can be computed securely. Thus, as long as strictly less than one third of the participants are malicious, regardless of their technological and computational power, they will not be able to learn information about the inputs of honest participants or force the protocol to output an incorrect result. Thereafter, we denote by $n$ the total number of participants involved in the secure computation and $t$ the number of malicious (\emph{i.e.}, corrupted) participants. Most of these early protocols for secure computation relies on multiplication and addition over a large field rather than on bits and NAND gates but they can be applied to compute any generic boolean circuit. In particular, these protocols are not efficient because each multiplication (\emph{i.e.}, AND gate) require interactions between all participants. 

Although the original techniques for secure computation are now approximately 25 years old, the field is still very active. The generic construction for secure computation requiring a strict majority of honest participants was originally presented in \cite{RB89} but that construction require a secure broadcast channel. Since then, more efficient solutions have been designed. The construction of \cite{BtH08} only requires linear communication complexity in the number of players for each multiplication gate but can only tolerate $t<n/3$ corrupt players. The construction of \cite{BFO12} requires approximately $O(n \log n)$ communication bits per multiplication gate and can tolerate any minority of malicious participants. With respect to the asynchronous setting, the construction proposed in \cite{CHP13} is secure as long as $t<n/3$, and has a communication complexity proportional to the number of players multiplied by the number of multiplication gates of the circuit representing the function to be computed. The work of \cite{CDI+13} can tolerate up to $n(1/2 - \epsilon)$ for any $\epsilon>0$ (a smaller value epsilon increases the size of the constant) of malicious players and achieves a linear communication complexity in the number of players per multiplication gate. Finally, other theoretically interesting protocols include \cite{DIKNS08,DIK10}.

\subsection{Secure Computation without Honest Majority}

In some situations, the assumption that two thirds or a majority of the participants are honest may be too strong. For instance, a government might be unwilling to share information if there is a risk that the other parties might leak some information (thus breaching the privacy property). It might be also the case that the information is so sensitive and trust so limited that participants only trust that a particular party will protect their data.  In particular, this means that even if up to $n-1$ participants are corrupted, the protocol is still secure. For this setting, it was shown in \cite{Cleve86} that certain distributed tasks are impossible without an honest majority. Thus it is impossible to achieve secure computation in full generality in this context without making either computational assumptions or physical assumptions such as the availability of trusted hardware.

A powerful and elegant solution is to base the construction of secure computation on a cryptographic primitive called \emph{oblivious transfer} \cite{Rabin81,EGL85,Crepeau88}. In this asymmetric bipartite primitive, the sender sends two messages in the oblivious transfer and the receiver gets to choose one of these values. In addition, the sender is oblivious to the choice of the receiver (\emph{i.e.}, he gets no information about this choice) and the receiver does not learn any information on the message he did not choose to learn. This primitive can be implemented using computational assumptions and is universal in the sense that the secure computation of any function can be implemented even if almost all participants are corrupted.

The protocol proposed in \cite{NNOB12} is based on \cite{CGT95} and provides secure computation solely from oblivious transfer. Both protocols first shares the inputs between two participants. Once this sharing has occurred the parties can perform operations such as AND, OR and XOR on distributed bits by relying on oblivious transfer. The latest construction only requires a constant number of bits of communication per gate. When extended to multiple participants, the complexity grows linearly in the number of players and the size of the circuit to be evaluated. Another construction of interest for secure computation is the IPS compiler named after its authors \cite{IPS08} (an optimization was presented in \cite{LOP11}). In these constructions, the participants imagine virtual players taking part in a protocol with honest majority evaluating the function in question. The participants will not just simulate these players but each player will verify by relying on oblivious transfer that a small minority of theses imagined players are acting honestly. As such, since making the protocol with honest majority fail would require that a participant corrupt a majority of virtual players, the honest participant will detect any effective cheating except with negligible probability. The communication complexity of \cite{LOP11} is proportional to the size of circuits multiplied by the number of players.

Finally, more recent protocols \cite{DPSZ12,DMKPSS13,KSS13,MS14,W14} allow the evaluation of circuits with a complexity that is linear in the size of the circuit and the number of players, with the aid of pre-processing using somewhat homomorphic encryption. These protocols rely on the idea of multiplicative triples in the information-theoretic setting.

\subsection{Applications}

In principle, the range of applications of secure computation is extremely large but relatively few implementations exist. The first public implementation was realized by Ivan Damg{\aa}rd and his team \cite{BCD+09}. This implementation was used in the context of an auction for the production rights of sugar beets. More precisely in Denmark, farmers can buy and sell production rights and the secure computation was used to find the market equilibrium price and the amount bought and sold for that price. The implementation consists of an honest majority protocol with three institutions. The event involved roughly 1200 farmers and took roughly half an hour. Since, the protocol is repeated annually. 

Secure computation can also used to perform general statistical data analysis. For instance, it was shown in \cite{DJ12}, that by using fully homomorphic encryption, private linear regression on $4 194 304$ elements can be completed in $256$ minutes.
Similarly, secure computation has also found applications in financial data analysis. In \cite{BTW12}, a solution was developed to analyze financial data in a private fashion for the ITL (Estonian Association of Information Technology and Telecommunications). This protocol is run roughly every six months. Another existing application of secure computation that was recently proposed is the detection of satellite collisions \cite{KW13}.

Secure computation can also protect the privacy of genetic and medical data \cite{KBLV13,EN13}. In terms of performance in \cite{KBLV13}, the authors have compared the secure version of an algorithm against the standard algorithms and showed that the private algorithm was roughly 720 times slower. However, it is expected that further advances in secure computation will significantly reduce this factor. In addition, for some computations for which the privacy aspect is of paramount importance, this factor may not be so detrimental. Open source implementations of secure computation can be found in \cite{MNPS04,BNP08,EFLL12}

\section{Proposal for a Scaled Down Trustworthy}
\label{sect_poc}
 
In order to demonstrate the practicality of the crypto-democracy, we propose in this section a preliminary implementation for a minimalistic Trustworthy. More precisely, we would like to use this significantly scaled down version of the Trustworthy as a proof-of concept to study the usefulness and feasibility of this type of initiative. In addition, another important contribution of such a project is to raise the awareness and interest of the security and privacy communities as well as the public at large. This implementation of the Trustworthy would use 5 independent universities as institutions. The main objectives of this implementation is to store in a secure and private manner the sensitive data belonging to users and to be able to conduct privacy-preserving data analysis for research in fields such as medicine and humanities. In particular, the Trustworthy can be used to provide meaningful non-trivial statistics on the data collected while ensuring that no single entity will be able to access this data in clear.

In this position paper, we only outline the main features of this implementation. However following the publication of paper, if we observe that the interest generated is sufficient, then we are planning to publish a technical report providing more details about the implementation.
 
\subsection{Choice of Institutions}

Relying on universities to play the role of institutions when implementing the Trustworthy seems to be a natural choice. Indeed, universities appear as a relevant choice because they are relatively independent and are trustworthy to the eye of the public. They also generally have the resources as well as expertise (in particular if they have a good security or cryptography research team) to guarantee the security and trust with respect to the software, hardware and network required by such an endeavor. Furthermore, we believe that it should not be difficult to find 5 universities spread around the world that would be sufficiently motivated to participate to this project.

Once the institutions have been selected, it will be essential to define a clear governance structure. For instance, no single university should be the sole master of the Trustworthy. Rather, we propose that the direction, decisions and computations that the trusted third will take should be decided by a simple majority structure in which each institution has exactly one vote. In addition, for specific data that have a higher degree of sensibility a stronger constraint such as unanimity across the institutions can be implemented. In this situation, no computation will be performed on the data unless all the institutions agree on this. In particular the Trustworthy cannot reveal information locked with an unanimity threshold with a majority vote but a majority vote would be enough to restrict the access to the data of a user or to discard it.

\subsection{Functionalities Provided by the Trustworthy}

In this section, we discuss in more details the data that will be secure by the Trustworthy and the functionalities that it will provide to users. For this first implementation, we believe that we should limit the functionalities offered to a few that will demonstrate the potential of the approach but can already be realized today with the current state of technology. However at the end of this section, we suggest other functionalities that are somewhat more advanced and could be added later. 
 
\begin{itemize}
\item \emph{Secure note}. Basically, this functionality simply corresponds to the possibility for the user to store, access and modify a text file in an online manner in the Trustworthy. This text file could be use to keep safe information such as password, passport number or credit card details. Typically, this file could be search by the user like a standard document and should not be too large.
\item \emph{Encrypted files}. The Trustworthy could be used to encrypt files chosen by the user and be responsible for protecting the secrecy of the key. For instance, the encrypted file could be stored on the hard drive of the user or on a cloud provider like Dropbox, Google drive or Microsoft drive. However, the key would be given to the Trustworthy, who would be responsible for decrypting the files if requested by the user.
\item {\em Private email.} Offering the functionality of a private email service is a natural service that could be implemented through the Trustworthy. This functionality would enable persons to exchange information in a truly private manner and by requesting that the email contains only text, this would not require a huge amount of storage.  
\item \emph{Private survey}.  This last functionality is the most complex but also the most interesting feature of our proposal. Basically, the Trustworthy would be responsible for conducting surveys, for instance in the context of a research project in medicine or humanities. The respondent of a survey would login and fill it privately. Afterward, the data of his particular survey will never be reconstructed and seen by anyone in clear but rather statistics could be computed on the aggregate data of the different respondents to the survey and then made public. To ensure that the information of a particular user is sufficiently hidden into the global data, a condition on the minimal number of responses could be predefined before any analysis can be conducted on the data. In addition, ideally the list of statistics that will be computed should be known by all beforehand and the output should be sanitized to ensure privacy.  
\end{itemize}

All the data stored on the Trustworthy should be secure according to some threshold, which might vary depending on the sensitivity of the data or the privacy expectations of users. For instance, in the case of the secure note, private email and encrypted files the user could decide by himself which threshold he want to use while for the survey this threshold should be fixed in advance and publicly known before the experiment starts (\emph{i.e.}, before respondents begin to answer to the survey). In order to reconstruct the private data or to compute on it, \emph{at least} the number of institutions required by the threshold have to cooperate together.

In practice, we envision that users will access to the Trustworthy in a transparent manner through their browser. Each user should have a unique identity and connect through a university of his choice using the corresponding software. This software should be open-source so that it can be inspected by anyone as well as being guaranteed and checked by the institutions. First when connecting, a mutual authentication protocol is run in which the user securely authenticates to the trusted third and vice-versa. 

\subsection{Implementing the Trustworthy}

For the functionalities that only need to perform secret sharing then the Shamir scheme will be used. All symmetric encryption should be made using a standard encryption scheme with long keys (\emph{e.g.}, AES with a $256$ bits key). The keys used should be created uniformly at random in order to ensure a strong level of security.   

\begin{itemize}
\item {\em Communication structure.} All the communications done between the user and the institutions are secured through the use of a public-key cryptosystem with a large key (possibly in conjunction with a protocol such as TLS) while the communications exchanged between the institutions are secure through the use of symmetric encryption. We also assume that the computer of the user is secure enough such that malware cannot intercept the information entered by the user in his web browser when connecting to the Trustworthy.
\item {\em Identification and authentication.} All the interactions between the Trustworthy and a user will be done through his browser. Several standard authentication techniques such as the use of a passphrase or two-way authentication based on certificates could be used but of course their applicability and security have to be evaluated in more details with respect to the architecture developed in the project. 
While the security provided by the authentication mechanism used is of paramount importance for ensuring the privacy and integrity of the users' data entrusted to the Trustworthy, a full discussion of the different alternatives for authentication as well as their pros and cons is out of the scope of this paper. Note however that if the adversary is able to impersonate a specific user during authentication, this only jeopardize the privacy of the data of this user but not of the data of all users.
\item {\em Secure note.}  The corresponding text file should be compressed, encrypted and then saved on the hard drive of each institution. The key will be stored in the Trustworthy using a secret sharing scheme with a threshold defined by the user. The user can view and modify his file online and securely through a web application running on his browser.
\item {\em Encrypted files.}  Similarly to the previous functionality, the Trustworthy would only be responsible to store the key needed to decrypt the encrypted files. The user can choose whether he want to store these files on a disk under his control or on a server of his choice. The Trustworthy offers the possibility to perform an online encryption and decryption of these files.
\item {\em Private email.} This functionality could be implemented by having the email sent a particular user stored in the Trustworthy using a secret sharing scheme such as that no institution would have a direct access to the content of these emails. In addition each email could itself be encrypted with the public key of the recipient. Only this recipient would be allowed to retrieve the corresponding shares once he has authenticated to the Trustworthy. As one of the assumptions is that the different institutions forming the Trustworthy are spread all around the world, the privacy of the emails would be ensured even if one of these institutions is requested to hand over the information it has (\emph{e.g.}, similarly to what happen to Lavabit and Silent Circle).
\item {\em Private survey.}  The private survey functionality can be implemented by combining the techniques described in \cite{BtH08} and \cite{BCD+09}. The survey can be represented in binary format as well as an integer to speed-up calculations. In a nutshell, the integer would be stored using a redundant representation (base 2, integer and base 1 for small integer) to simplify the computations. Data consistency between the binary and the integer representation will be verified by the Trustworthy. Examples of possible operations on the survey data include AND and NOT (for the binary representation), the sum and multiplication (for integer representation) and the transfer from binary to integer (or vice-versa). It is also straightforward to compute if an encrypted value is smaller or larger than a particular threshold. The implementation that we propose is powerful enough to compute interesting non-trivial statistics. For instance, we could easily compute the percentage of female participants whose age is between $32$ and $40$ years with both diabetes and the Coeliac disease. More precisely, this statistic requires the computation of $2$ inequalities and $3$ multiplications for each user in the survey. However, as the scalability increases with the number of users, this computation could be performed with only a constant number of messages sent. 
\end{itemize}

The cost per institution will be quite low as most of the infrastructure is already provided by the university. With respect to the additional hardware needed, we believe that a 3000\$ server under the responsibility of each institution should be sufficient to start the project. In particular, since the Trustworthy is mainly responsible of the storing of the encryption keys (at least for the secure notes and the encrypted files functionalities), the memory requirements are relatively small. By far, the main investment for the practical realization of this project will be the creation of the appropriate software although indirect but limited costs have also to be taken into account (\emph{e.g.}, for the maintenance and possible upgrades). 
We believe that a two-year period for conducting this project is realistic.  

Other \emph{internal} functionalities need to be added to the Trustworthy. For instance, it should be possible to reshare information with a new threshold, which corresponds in practice to transforming the data encoded from one secret sharing to another. In case of a hardware malfunction, it should be possible to retrieve the corresponding data. Finally, it should also be possible to permanently remove data from the servers (and the backup) if a user request it, thus implementing a form of right to be forgotten. 
This functionality could be implemented for instance by erasing the corresponding shares at the level of each institution upon the user's request.

\section{Conclusion}
\label{sect_conclu}

Our main objective in this paper was to show that the field of secure computation can help to the improvement of democracy. In particular, we have highlighted how using current technologies, we could implement the Trustworthy, which is a ideal trusted third party that could manage our personal data. The objective of the Trustworthy is to ensure a high degree of actual and perceived privacy while simplifying our lives. In order to show the feasibility of the approach, we have also propose a realistic project, achievable with limited financial resources that is interesting and powerful enough to generate interest and to enable researchers to experiment on this concept. 

In the future, we would like to investigate also additional functionalities that could be added to the project once the basic functionalities described previously have been implemented. For instance, we hope that the system will be powerful and fast enough to perform more complex functions on the private survey data such as advanced data mining or machine learning algorithms.

\section*{Acknowledgments} 

S\'ebastien is supported by the Inria large scale project CAPPRIS (Collaborative Action on the Protection of Privacy Rights). Samuel is supported by the European Research Commission Starting Grant 279447, 
the Danish National Research Foundation and The National Science Foundation of China for the Sino-Danish Center for the Theory of Interactive Computation. 

\bibliographystyle{alpha} 
\bibliography{references}        

\end{document}